\documentclass[10pt,showpacs,preprintnumbers,amsmath,amssymb,
aps,prd,nofootinbib,eqsecnum,a4paper]{revtex4}
\usepackage{graphicx,epsf}
\def\be{\begin{equation}}
\def\ee{\end{equation}}
\def\bea{\begin{eqnarray}}
\def\eea{\end{eqnarray}}
\def\p{\partial}

\def\Q{{\cal{Q}}}
\def\vp{{\varphi}}
\newcommand\dvp[1]{{\delta\varphi_{#1}}}
\newcommand\dvpI[1]{{\delta\varphi_{{#1}I}}}
\newcommand\dvpK[1]{{\delta\varphi_{{#1}K}}}
\newcommand\dvpL[1]{{\delta\varphi_{{#1}L}}}

\newcommand\dU[1]{{\delta U_{#1}}}
\def\H{{\cal H}}
\def\cs2{c_{\rm{s}}^2}
\def\U0{{\bar U_0}}
\def\wt{\widetilde}
\def\dT{{\delta{\bf T}_1}}
\def\dTT{{\delta{\bf T}_2}}

\def\12{\frac{1}{2}}
\def\BkBk{{B_{1,k}B_{1,}^{~k}}}
\def\ppij{{\p^{-1}_i\p^{-1}_j}}
\def\dvpdvpKll{\delta\vp_{1K,l}\delta\vp_{1K,}^{~~~~l}}
\def\Xkdvk{{\sum_K X_K \delta\vp_{1K}}}
\def\M{{\cal{M}}}
\def\k{{\bf{k}}}

\def\kvi{{{k^i}}}
\def\qvi{{{q^i}}}
\def\pvi{{{p^i}}}
\newcommand\eq[1]{Eq.~(\ref{#1})}
\newcommand\eqs[1]{Eqs.~(\ref{#1})}
\begin{document}
\preprint{} 
\title{A not so short note on the Klein-Gordon equation at
second order} 
\author{Karim A.~Malik} 
\affiliation{Cosmology and Astroparticle Physics Group, Department of
Physics, University of Lancaster, Lancaster LA1 4YB, 
United Kingdom\footnote{Present Address: 
Astronomy Unit, School of Mathematical Sciences, Queen Mary University
of London, Mile End Road, London, E1 4NS, United Kingdom} }
\date{\today}
\begin{abstract}
We give the governing equations for multiple scalar fields in a flat
Friedmann-Robertson-Walker (FRW) background spacetime on all scales,
allowing for metric and field perturbations up to second order. We
then derive the Klein-Gordon equation at second order in closed form
in terms of gauge-invariant perturbations of the fields in the uniform
curvature gauge. We also give a simplified form of the Klein-Gordon
equation using the slow-roll approximation.
\end{abstract}

\pacs{98.80.Cq \hfill  JCAP03(2007)004, astro-ph/0610864v5}

\maketitle

\section{Introduction}

In the cosmological standard model the universe undergoes at very
early times a period of inflation, during which fluctuations in the
scalar fields are stretched to super-horizon scales and later on
source the Cosmic Microwave Background (CMB) anisotropies and the
large scale structure.
In standard inflation models the scalar field(s) responsible for the
accelerated expansion of the universe also provide(s) these
fluctuations \cite{LLBook}.
If we are interested in calculating observational consequences of the
scalar field dynamics, linear or first order cosmological perturbation
theory is sufficient, if our main interest is in calculating the power
spectrum of the fluctuations.
However, if we are interested in higher order observables, such as the
bispectrum, we also need cosmological perturbation theory beyond
linear order. The study of higher order effects promises to give new
insights into the physics of the early universe. Much progress has
been made recently to provide the necessary theoretical tools to study
inflation beyond the standard linear approximation. In this article we
use second order perturbation theory to derive the Klein-Gordon
equation governing the evolution of the scalar fields during and after
inflation.


At the moment there are mainly two approaches to study higher order
effects and in particular non-gaussianity: 
One approach uses nonlinear theory and the separate universe
approximation, either employing a gradient expansion \cite{SB,RS} or
using the $\Delta N$ formalism \cite{SaSt95,SaTa,LMS,Lyth:2005fi,LV}
(see also \cite{Starobinsky} for early works).
%
Both the gradient expansion and the $\Delta N$ formalism are so far
only usable on large scales, that is scales much larger than the
particle horizon, which is sufficient for many applications.
The other approach uses second order cosmological perturbation theory
following Bardeen
\cite{
Bardeen80,Mukhanov,Bruni,Acquaviva,Maldacena,Nakamura,Noh,Bartolo:2001cw,
Rigopoulos:2002mc,Bernardeau,MW2004,BartoloJCAP,Finelli:2003bp,
Bartolo:2004if,Enqvist:2004bk,filippo,Tomita:2005et,
Lyth:2005du,Seery2005,M2005,Seery:2006}.
The Bardeen approach is valid on all scales, it can be however more
complex than the $\Delta N$ formalism. Both approaches give the same
results on large scales.\\

In this article we focus on deriving the Klein-Gordon equation at
second order valid on all scales for multiple scalar fields minimally
coupled to gravity in a flat Friedmann-Robertson-Walker (FRW)
background spacetime. We take into account metric perturbations up to
second order, considering scalar perturbations only. Throughout this
article we shall work in the flat gauge.
Having derived the field equations up to second order in the
perturbations, we calculate the Klein-Gordon equation at second order
in closed form for multiple fields. This allows to show that the
Klein-Gordon equation also at this order does not contain terms that
diverge in the large scale limit. 
These divergences might have appeared through the introduction of
inverse gradients, but the calculations show that any inverse gradient
present is balanced by one or more gradient terms.  \\

The paper is organised as follows: in the next section we give the
governing equations, that is Einstein's field equations and the
Klein-Gordon equation up to second order in the flat gauge.
In Section \ref{kg_sec_real} we give the Klein-Gordon equation at
first and second order in closed form, that is solely in terms of the
field fluctuations in the flat gauge, in real space.
In Section \ref{kg_sec_fourier} we rewrite the second order
Klein-Gordon equation in Fourier space. 
In Section \ref{sr_sec} we employ the slow roll approximation to
simplify the Klein-Gordon equation in Fourier space.
We conclude in the final section.

Throughout this paper we assume a spatially flat
FRW background spacetime and use conformal time, $\eta$. Derivatives
with respect to conformal time are denoted by a dash. Greek indices,
$\mu,\nu,\lambda$, run from $0,\ldots 3$, while lower case Latin
indices, $i,j,k$, run from $1,\ldots3$. Upper case Latin indices,
$I,J,K$, denote different scalar fields.

\section{Governing equations}
\label{governing_sect}

In this section we briefly review the governing equations of a system
of multiple scalar fields. A more detailed exposition can be found for
example in Ref.~\cite{M2005}.
The covariant Einstein equations are given by
\be
\label{Einstein}
G_{\mu\nu}=8\pi G \; T_{\mu\nu} \,,
\ee
where $G_{\mu\nu}$ is the Einstein tensor, $T_{\mu\nu}$ the
total energy-momentum tensor, and $G$ Newton's constant.
Through the Bianchi identities, the field equations (\ref{Einstein})
give the local conservation of the total energy and momentum,
\be
\label{nablaTmunu}
\nabla_\mu T^{\mu\nu}=0\,,
\ee
where $\nabla_\mu$ is the covariant derivative.
The energy momentum tensor for $N$ scalar fields minimally coupled to
gravity is
\be
\label{multiTmunu}
T_{\mu\nu}=\sum_{K=1}^N\left[
\vp_{K,\mu}\vp_{K,\nu}
-\frac{1}{2}g_{\mu\nu}g^{\alpha\beta}\vp_{K,\alpha}\vp_{K,\beta}\right]
-g_{\mu\nu}U(\vp_1,\ldots,\vp_N)\,, 
\ee
where $\vp_K$ is the $K$th scalar field and $U$ the scalar field
potential and $\vp_{K,\mu}\equiv\frac{\p\vp_{K}}{\p x^\mu}$.\\

Tensorial quantities are separated into a background value and
perturbations according to
\be
\label{tensor_split}
{\bf T}= {\bf T}_0+\dT+\frac{1}{2}\dTT+\ldots\,,
\ee
where the background part is a time dependent quantity only ${\bf
T}_0\equiv {\bf T}_0(\eta)$, whereas the perturbations depend on time
and space coordinates $x^\mu=[\eta,x^i]$, that is ${\bf \delta
T}_n\equiv{\bf \delta T}_n(x^\mu)$. The order of the perturbation is
indicated by a subscript, i.e.~${\bf \delta T}_n =O(\epsilon^n)$,
$\epsilon$ here being the small parameter. \\

The metric tensor up to second order, including only scalar
perturbations, and without gauge constriction is
\bea
\label{metric1}
g_{00}&=&-a^2\left(1+2\phi_1+\phi_2\right) \,, \\
g_{0i}&=&a^2\left(B_1+\frac{1}{2}B_2\right)_{,i}\,, \\
g_{ij}&=&a^2\left[\left(1-2\psi_1-\psi_2\right)\delta_{ij}
+2E_{1,ij}+E_{2,ij}\right]\,,
\eea
where $a=a(\eta)$ is the scale factor, $\eta$ conformal time,
$\delta_{ij}$ is the flat background metric, $\phi_1$ and $\phi_2$ the
lapse functions, and $\psi_1$ and $\psi_2$ the curvature perturbations
at first and second order; $B_1$ and $B_2$ and $E_1$ and $E_2$ are
scalar perturbations describing the shear.
The contravariant form of the metric tensor is given in Appendix
\ref{app_met}.
We shall employ the flat slicing and threading throughout this
paper. In this gauge the spatial 3-hypersurfaces are flat and
we have 
\be
\label{defgauge}
\wt\psi_1=\wt\psi_2=\wt E_1=\wt E_2=0\,.
\ee

Details on how to construct gauge-invariant variables at second order
in general and in the flat gauge in particular can be found in
Refs.~\cite{MW2004} and \cite{M2005}. We simply state the result here.
%

The field fluctuation on uniform curvature hypersurfaces at first
order, also known as Sasaki-Mukhanov variable
\cite{Sasaki1986,Mukhanov88}, is given for the $I$th field as
\be
\label{defQ1I}
\wt{\dvpI1}=\dvpI1+\frac{\vp_{0I}'}{\H}\psi_1\,.
\ee
The field fluctuation in the flat gauge, or Sasaki-Mukhanov variable,
at second order for the $I$th field is \cite{MW2004,M2005}\footnote{
Rewritten using results from K.~A.~Malik and
D.~R.~Matravers, arXiv:0804.3276 [astro-ph].}
%
%
\be
\label{defQ2I}
\wt{\dvpI2}=\dvpI2+\frac{\vp_{0I}'}{\H}\psi_2
+\left(\frac{\psi_1}{\H}\right)^2\left[
2\H\vp_{0I}'+\vp_{0I}''-\frac{\H'}{\H}\vp_{0I}'\right]
+2\frac{\vp_{0I}'}{\H^2}\psi_1'\psi_1+\frac{2}{\H}\psi_1\dvpI1'
-2\delta\vp_{1,k}E_{1,}^{~k}
+{\cal{X}}\left(\psi,E\right)\,,
\ee
where ${\cal{X}}\left(\psi,E\right)$ contains terms quadratic in
gradients of the metric perturbations $\psi_1$ and $E_1$.
%
%
Since we are working in the flat gauge throughout this paper, we shall
drop the ``tilde'' to denote variables that have to be evaluated on
this gauge.  In particular, we denote the gauge invariant field
fluctuations in the flat gauge simply by $\dvp1$ and $\dvp2$, hoping
that this notation will make the equations more intuitive and
transparent. Note however, that the field fluctuations in the flat
gauge were denoted $\Q_1$ and $\Q_2$ at first and second order,
respectively, in Ref.~\cite{M2005}.

\subsection{Klein-Gordon equation}
\label{KG_sect}

We split the scalar fields $\vp_I$ into a background and perturbations
up to and including second order according to \eq{tensor_split},
\be
\vp_I(x^\mu)=\vp_{0I}(\eta)+\dvp{1I}(x^\mu)+\frac{1}{2}\dvp{2I}(x^\mu)\,.
\ee
The potential $U\equiv U(\vp_I)$ can be split similarly according to
\be
U(\vp_I)=U_0+\dU1+\frac{1}{2}\dU2\,,
\ee
where
%
%
%
%
\bea
\label{defdU1}
\dU1&=&\sum_K U_{,\vp_K}  \dvp{1K}\,,\\
\label{defdU2}
\dU2&=&\sum_{K,L}U_{,\vp_K\vp_L}\dvp{1K}\dvp{1L}
+\sum_K U_{,\vp_K}\dvp{2K}\,.
\eea
and we use the shorthand $U_{,\vp_K}\equiv\frac{\p U}{\p\vp_K}$.
The energy-momentum tensor, \eq{multiTmunu}, expanded up to second
order in the perturbations for the metric tensor \eq{metric1} is given
in Appendix \ref{Tmunu_sect} (in the flat gauge).\\

The evolution of the scalar fields is governed by the Klein-Gordon
equation, which is the energy conservation equation,
(\ref{nablaTmunu}), for the energy momentum tensor defined in
\eq{multiTmunu}.
Using the energy momentum tensor given in Appendix \ref{Tmunu_sect}
and the energy conservation equation (\ref{nablaTmunu}) we get the
Klein-Gordon equation in the multi field case at zeroth order for the
$I$th field
\bea
\label{KGback}
\vp_{0I}''+2\H\vp_{0I}'+a^2 U_{,\vp_I}=0\,,
\eea
where $\H\equiv\frac{a'}{a}$, at first order
\be
\label{KG1flatsingle}
\dvpI1''+2\H\dvpI1'+2a^2 U_{,\vp_I}\phi_1
-\nabla^2\dvpI1-\vp_{0I}'\nabla^2 B_1
-\vp_{0I}'\phi'_1+a^2 \sum_K U_{,\vp_I\vp_K}\dvpK1
=0\,,
\ee
and we finally get at second order
\bea
\label{KG2flatsingle}
\dvpI2''&+&2\H\dvpI2'-\nabla^2\dvpI2+a^2 \sum_K U_{,\vp_I\vp_K}\dvpK2
+a^2 \sum_{K,L} U_{,\vp_I\vp_K\vp_L} \dvpK1\dvpL1 +2a^2 U_{,\vp_I}\phi_2
-\vp_{0I}'\left(\nabla^2 B_2+\phi_2'\right)\nonumber\\
&+&4\vp_{0I}' B_{1,k}\phi_{1,}^{~k}
+2\left(2\H\vp_{0I}'+a^2 U_{,\vp_I}\right) B_{1,k}B_{1,}^{~k}
+4\phi_1\left(a^2\sum_K U_{,\vp_I\vp_K}\dvpK1-\nabla^2\dvpI1\right)
+4\vp_{0I}'\phi_1\phi_1'\nonumber\\
&-&2\dvpI1'\left(\nabla^2 B_1+\phi_1'\right)-4\dvpI1'_{,k}B_{1,}^{~k}
=0\,,
\eea
where all the variables are in the flat gauge, as stressed in the
previous section.

\subsection{The field equations}
\label{field_sect}

In this section we give the field equations as given from
\eq{Einstein} in the background, at first and at second order in the
perturbations. Again we would like to stress that the perturbed
equations are in the flat gauge throughout this section.

At zeroth order we get the Friedmann equation from the $0-0$
component of the Einstein equations
\be
\label{Friedmann}
\H^2=\frac{8\pi G}{3}  
\left(\sum_K\frac{1}{2}{\vp'_{0K}}^2+a^2 U_0\right)  \,,
\ee
and from the $i-j$ component we find
%
%
\be
\label{ij_back}
\left(\frac{a'}{a}\right)^2-2\frac{a''}{a}=8\pi G 
\left(\sum_K\frac{1}{2}{\vp'_{0K}}^2-a^2 U_0\right)  \,.
\ee

\subsubsection{First order}
\label{field1}

Using the background equations given above, we find the $0-0$
component of the Einstein equations at first order to be
\be
\label{00Ein1}
2a^2U_0{\phi_1}+\sum_K\vp_{0K}'{{\dvpK1}}'+a^2{\delta U_1}
+\frac{\H}{4\pi G}\nabla^2B_1
=0\,,
\ee
the $0-i$ part gives
\be
\label{0iEin1}
\H\phi_1-4\pi G\sum_K\vp_{0K}'{{\dvpK1}}=0\,.
\ee
{}From the $i-j$ component of the Einstein equation we get the trace free part
\be
\label{offtrace1}
B_1'+2\H B_1+\phi_1=0\,.
\ee
Using \eq{offtrace1} and the background field equations, we get the
first order trace in its simplest form
\be
\label{trace1}
\H\phi_1'+4\pi G\left[ 
a^2\delta U_1+2a^2U_0{\phi_1}-\sum_K\vp_{0K}'{{\dvpK1}}'\right]
=0\,.
\ee
%

\subsubsection{Second order}
\label{field2}

The $0-0$ Einstein equation at second order, using the first and
zeroth order $0-0$ equations, \eqs{Friedmann} and (\ref{00Ein1}), is
given by
%
%
\bea
\label{Ein00_2}
8\pi G a^2 U_0\left(\phi_2+\BkBk\right)
&+&\H\nabla^2B_2 
+\12\left[B_{1,kl}B_{1,}^{~kl}
-\left(\nabla^2 B_1\right)^2\right]
-2\H \phi_{1,k}B_{1,}^{~k}\nonumber\\
&&+4 \pi G\sum_K\left[
\vp_{0K}'\dvpK2' +a^2\delta U_2+4 a^2\delta U_1\phi_1
+{\dvpK1'}^2+\dvpK1_{,k}\dvpK1_{,}^{~k}
\right]=0\,.
\eea
The $0-i$ Einstein equation is given by
\be
\label{0i_2}
\H\phi_{2,i}-4\H\phi_{1}\phi_{1,i}+2\H B_{1,ki}B_{1,}^{~k}
+B_{1,ki}\phi_{1,}^{~k}-\nabla^2 B_1\phi_{1,i}
-4\pi G \sum_K\left[\vp_{0K}'\dvpK2_{,i}+2\dvpK1'\dvpK1_{,i}\right]=0\,.
\ee
We then use the first order $0-i$ equation, \eq{0iEin1}, rewrite
\eq{0i_2} and take the trace, which gives
%
%
\bea
\label{0i_2version2}
&&\H\left(\phi_{2}-2\phi_{1}^2+ B_{1,k}B_{1,}^{~k}\right)
-4\pi G \sum_K\vp_{0K}'\dvpK2\nonumber\\
&&\qquad
+\nabla^{-2}\left(\phi_{1,kl}B_{1,}^{~~kl}
-\nabla^2 B_1\nabla^2\phi_1\right)
-8\pi G \sum_K \nabla^{-2}\left(
\dvpK1'\nabla^{2}\dvpK1+\delta\vp_{1K,l}'\delta\vp_{1K,}^{~~~~l}
\right)
=0\,,
\eea
where we introduce the inverse Laplacian,
$\nabla^{-2}(\nabla^{2})X=X$.
%
%
%
The $i-j$ Einstein equation is given by
\bea
\label{ij2}
&&\Bigg\{
16\pi G a^2 U_0\left(\phi_2-4\phi_1^2+\BkBk\right)
+2\H\phi_2'-8\H\phi_1\phi_1'-2\phi_{1,k}\phi_{1,}^{~k}
+4\H B_{1,k}'B_{1,}^{~~k}\nonumber\\
&&\qquad
+\nabla^2\left(B_2'+2\H B_2+\phi_2\right)-2\phi_1'\nabla^2 B_1
+B_{1,kl}B_{1,}^{~kl}-\left(\nabla^2 B_1\right)^2
+32\pi G\left(a^2 \delta U_1+2 a^2 U_0\phi_1\right)\phi_1\nonumber\\
&&\qquad
+8\pi G\left[ 
\sum_K\left(
\dvpK1_{,l}\dvpK1_{,}^{~~l}-\vp_{0K}'\dvpK2'-{\dvpK1'}^2\right)
+a^2\delta U_2\right]\Bigg\}\delta^i_{~j}\\
&&\qquad
-\left(B_2'+2\H B_2+\phi_2\right)^{~i}_{,~j}
+2\phi_{1,}^{~~i}\,\phi_{1,j}+2B_{1,~j}^{~~i}\left(\phi_1'+\nabla^2 B_1\right)
-2B_{1,~k}^{~~i}B_{1,~j}^{~~k}
-16\pi G\sum_K\dvpK1_{,}^{~i}\dvpK1_{,j}=0\,.\nonumber
\eea
Taking now the divergence of \eq{ij2} twice\footnote{Following on
the same lines as detailed in K.~A.~Malik, D.~Seery and K.~N.~Ananda,
arXiv:0712.1787 [astro-ph].} we get
\bea
\label{divdivij2}
&&
8\pi G a^2 U_0\left(\phi_2-4\phi_1^2+\BkBk\right)
+\H\phi_2'-4\H\phi_1\phi_1'-\phi_{1,k}\phi_{1,}^{~k}
+2\H B_{1,k}'B_{1,}^{~~k}
+\frac{1}{2}\left(B_{1,kl}B_{1,}^{~kl}-\left(\nabla^2 B_1\right)^2\right)
\nonumber\\
&&
-\phi_1'\nabla^2 B_1+16\pi G\left(a^2 \delta U_1+2 a^2 U_0\phi_1\right)\phi_1
+4\pi G\left[ 
\sum_K\left(
\dvpK1_{,l}\dvpK1_{,}^{~~l}-\vp_{0K}'\dvpK2'-{\dvpK1'}^2\right)
+a^2\delta U_2\right]\nonumber\\
&&
+\nabla^{-2}\Big\{
\phi_{1,}^{~~i}\,\phi_{1,j}+B_{1,~j}^{~~i}\left(\phi_1'+\nabla^2 B_1\right)
-B_{1,~k}^{~~i}B_{1,~j}^{~~k}
-8\pi G\sum_K\dvpK1_{,}^{~i}\dvpK1_{,j}\Big\}^{~j}_{,i}=0\,,
\eea
where we made use of the relation 
$
\p_i\p^j\left[
\left(f_2+f_1g_1\right)\delta^i_{~j}
\right]
=\nabla^2\left(f_2+f_1g_1\right)
=\nabla^2f_2+f_1\nabla^2 g_1+g_1\nabla^2 f_1+2f_{1,k}g_{1,}^{~k}  
$.
The spatial trace is finally given by contracting \eq{ij2} as 
%
%
\bea
\label{spatialtrace2}
&&24\pi G a^2 U_0\left(\phi_2-4\phi_1^2+\BkBk\right)
-2\phi_{1,k}\phi_{1,}^{~k}
-2\phi_1'\nabla^2B_1
-6\H B_{1,}^{~k} \left(2\H B_{1,k}+\phi_{1,k}\right)\nonumber\\
&&+3\H\phi_2'-12\H\phi_1\phi_1'
+\nabla^2\left(B_2'+2\H B_2+\phi_2\right)
+\frac{1}{2}\left[B_{1,kl}B_{1,}^{~kl}-\left(\nabla^2 B_1\right)^2\right]
+48\pi G\phi_1\left(a^2 \delta U_1+2 a^2 U_0\phi_1\right)
\nonumber\\
&&+4 \pi G\sum_K\left( 3a^2\delta U_2-3\vp_{0K}'\dvpK2'-3{\dvpK1'}^2
+\dvpK1_{,l}\dvpK1_{,}^{~~l}\right)
=0\,.
\eea
In the above we made copious use of the background field equations
without explicitly stating it.
%

\section{The Klein Gordon equation in closed form}
\label{kg_sec_real}

Using the field equations given in the previous section we can now
rewrite the Klein Gordon equations at first and second order in the
multiple field case.
At first order the Klein-Gordon equation is, from \eq{KG1flatsingle},
\be
\label{KG1_flat}
\dvpI1''+2\H\dvpI1'-\nabla^2\dvpI1
+a^2\sum_K\left\{
U_{,\vp_K\vp_I}
+\frac{8 \pi G}{\H}\left(
\vp_{0I}'U_{,\vp_K}+\vp_{0K}'U_{,\vp_I}
+\vp_{0K}'\vp_{0I}'\frac{8 \pi G}{\H}U_0
\right)
\right\}\dvpK1=0\,.
\ee
Although there are notationally more compact forms of the Klein-Gordon
equation at first order \cite{Hwang,Taruya}, we find this form
particularly easy to use. All that is required once a model is chosen
is to calculate the derivatives of the potential and no further
calculation is necessary.

It will be useful in the following to define
\be
\label{defXI}
X_I\equiv a^2\left(
\frac{8\pi G}{\H}U_0\vp_{0I}'+U_{,\vp_I}
\right)\,.
\ee
We finally get the Klein-Gordon equation for multiple fields in the
flat gauge, from \eq{KG2flatsingle}, and using all the field equations
given in the previous section
%
%
%
\bea
\label{flatKG2real}
\dvpI2''&+&2\H\dvpI2'-\nabla^2\dvpI2
+a^2\sum_K\left[
U_{,\vp_K\vp_I}+\frac{8 \pi G}{\H}\left(
\vp_{0I}'U_{,\vp_K}+\vp_{0K}'U_{,\vp_I}
+\vp_{0K}'\vp_{0I}'\frac{8 \pi G}{\H}U_0
\right)
\right]\dvpK2 \nonumber\\
&+&\frac{16\pi G}{\H}\Bigg[\
 \dvpI1' \sum_K X_K\dvpK1
+\sum_K\vp_{0K}'\dvpK1 \sum_K a^2 U_{,\vp_I\vp_K}\dvpK1
\Bigg] \nonumber\\
&+&\left(\frac{8\pi G}{\H}\right)^2 \sum_K\vp_{0K}'\dvpK1
\Bigg[\
a^2 U_{,\vp_I}\sum_K\vp_{0K}'\dvpK1
+\vp_{0I}'
\sum_K\left(a^2 U_{,\vp_K}+X_K\right)\dvpK1
\Bigg]\nonumber\\
&-&2\left(\frac{4\pi G}{\H}\right)^2\frac{\vp_{0I}'}{\H}
\sum_K X_K\dvpK1 \sum_K \left( X_K\dvpK1
+ \vp_{0K}'\dvpK1'\right)
+\frac{4\pi G}{\H}\vp_{0I}'\sum_K{\dvpK1'}^2
\nonumber\\
&+&a^2\sum_{K,L} \left[
U_{,\vp_I\vp_K\vp_L} 
+ \frac{8\pi G}{\H}\vp_{0I}' U_{,\vp_K\vp_L}
\right]\dvpK1\dvpL1
+F\left(\dvpK1',\dvpK1\right)=0\,,
\eea
where $F\left(\dvpK1',\dvpK1\right)$ contains gradients and inverse
gradients quadratic in the field fluctuations and is defined as
%
%
\bea
\label{Fdvk1}
&&\hspace{-5mm}
F\left(\dvpK1',\dvpK1\right)
%
=
\left(\frac{8\pi G}{\H}\right)^2
\delta\vp_{1I,l}'\nabla^{-2}\sum_K\left(
X_K\dvpK1+\vp_{0K}'\dvpK1'\right)_{,}^{~l}
-\frac{16\pi G}{\H}\nabla^2\dvpI1\sum_K\vp_{0K}'\dvpK1
\nonumber\\
&&\hspace{-5mm}
+2\frac{X_I}{\H}\left(\frac{4\pi G}{\H}\right)^2 
\nabla^{-2}
\Bigg[
\sum_K\vp_{0K}'\delta\vp_{1K,lm}
\nabla^{-2}\sum_K\left(X_K\dvpK1+\vp_{0K}'\dvpK1'\right)_{,}^{~lm}
-\sum_K\left(X_K\dvpK1+\vp_{0K}'\dvpK1'\right)
\nabla^{2}\sum_K\vp_{0K}'\dvpK1
\Bigg]
\nonumber\\
&&\hspace{-5mm}
+\frac{4\pi G}{\H}
\Bigg[\vp_{0I}'\sum_K \delta\vp_{1K,l}\delta\vp_{1K,}^{~~~~l}
+4X_I\nabla^{-2}\sum_K\left(
\dvpK1'\nabla^2\dvpK1+\dvpK1'_{,l}\dvpK1_{,}^{~l}\right)
\Bigg]
\nonumber\\
&&\hspace{-5mm}
+\left(\frac{4\pi G}{\H}\right)^2
\frac{\vp_{0I}'}{\H}\Bigg[
\nabla^{-2}\sum_K\left( X_K\dvpK1+\vp_{0K}'\dvpK1'\right)_{,lm}
\nabla^{-2}\sum_K\left(X_K\dvpK1+\vp_{0K}'\dvpK1'\right)_{,}^{~lm}
-\sum_K\vp_{0K}'\delta\vp_{1K,l}
\sum_K\vp_{0K}'\delta\vp_{1K,}^{~~~~l}
\Bigg]\nonumber\\
&&\hspace{-5mm}
-\frac{\vp_{0I}'}{\H} \nabla^{-2}
\Bigg\{
8\pi G\sum_K\left(\dvpK1_{,l}\nabla^{2}\dvpK1_{,}^{~l}
+\nabla^{2}\dvpK1\nabla^{2}\dvpK1
+\dvpK1'\nabla^{2}\dvpK1'+\dvpK1'_{,l}\dvpK1_{,}^{\prime~l}\right)
\nonumber\\
&&\qquad\qquad
-\left(\frac{4\pi G}{\H}\right)^2
\Bigg[
2\nabla^{-2}\sum_K\left(X_K\dvpK1+\vp_{0K}'\dvpK1'\right)_{,~j}^{~i}
\sum_K X_K\dvpK1
+\sum_K\vp_{0K}'\dvpK1_{,}^{~i}\sum_K\vp_{0K}'\dvpK1_{,j}
\Bigg]_{,i}^{~j}
\Bigg\}\,.
\eea
This is the Klein-Gordon equation at second order for multiple scalar
fields in terms of the field fluctuations $\dvpI2$ and $\dvpI1$, at
second and first order, respectively, in the flat gauge. No slow roll
is assumed.

Note that terms containing inverse gradients and inverse Laplacians
\emph{never} appear on their own. They are always either multiplied by
terms containing gradients and Laplacians or contain these
themselves. Moreover, the order of the ``positive'' gradients is
either the same or higher as that of the ``negative'' or inverse
gradients. Hence the large scale limit stays well defined and no
infrared cut-off has to be introduced by hand.\\

As an intermediate step in the derivation of the Klein-Gordon equation
above, the equation is given in Appendix \ref{KG2_intermediate} in
terms of the field fluctuations as in \eq{flatKG2real}, but also with
metric potentials $\phi_1$ and $B_1$, which allows for a more compact
presentation.
Using \eqs{00Ein1} and (\ref{0iEin1}) for $\phi_1$ and $B_1$,
respectively, the intermediate Klein-Gordon equation,
\eq{flatKG2_intermediate}, can readily be solved numerically either in
real space or in Fourier space (as outlined in the following
section). However, for analytical studies \eq{flatKG2real} above seems
preferable.

\section{The Klein Gordon equation in closed form in Fourier space}
\label{kg_sec_fourier}

In order to solve the Klein-Gordon equation it has become standard
procedure to work in Fourier space instead of real space (see
e.g.~Ref.~\cite{LLBook}). However, at second order things are slightly
more complicated than at first order, where the shift to Fourier
components can be nearly done implicitly, and we shall look at these
complications below.

A perturbation $\delta\vp$ is related to its Fourier component
$\delta\vp_\k$ as
\be 
\delta \vp(\eta,x^i)
= \int d^3 k \delta\vp_\k \exp\left(i k_i x^i\right) \,,
\ee
where $\delta \vp_\k=\delta \vp_\k(\eta,k^i)$, and $k^i$ is the comoving
wavenumber. 
We follow the notation of Ref.~\cite{LLBook} and use in the following
the same symbol for the real space and the Fourier space variable,
discriminating the two by the respective argument, i.e.~$\delta
\vp(k^i)$ instead of $\delta \vp_{\k}$, whenever convenient.\\

At first order the procedure to transform the Klein-Gordon equation is
straight forward: the only change to the equation is to replace
$\dvpI1$ by $\dvpI1(k^i)$ in \eq{KG1_flat} and the Laplacian by
\be
\label{lap}
\nabla^{2} \to -k^{2} \,,
\ee
where $k^2\equiv k^i k_i$.
At second order things are slightly more complicated, as already
pointed out above.
Second order variables proper, like $\dvp2$, follow the same rules as
at first order and can simply be replaced. However the terms quadratic
in the first order variables, like $\dvp1^2$, have now to be treated
using the convolution theorem, which is given for two functions $f$
and $g$, both functions of the coordinates $x^\mu$, by
\be
\label{convolution}
\left[f\ g\right]_\k
= \int d^3 \lambda\  f(p^i) g(q^i)\,,
\ee
where we introduced 
%
$d^3 \lambda\equiv\frac{d^3pd^3q}{(2\pi)^3}\delta^3(k^i-p^i-q^i)$.


In addition to the Laplace-operator we now also have gradients and
inverse gradient operators, which in Fourier space translate into
\be
\label{gradients}
\p_i \to -i k_i\,, \qquad\p_i^{-1} \to i \frac{k^i}{k^2}\,,
\ee
and similarly the inverse Laplacian 
\be
\label{inv_lap}
\nabla^{-2} \to -k^{-2}\,.
\ee

We have now assembled all the necessary tools to rewrite the
Klein-Gordon equation at second order in terms of its Fourier
components.
Rewriting the first part which is independent of gradients is simple
and only necessitates in \eq{flatKG2real} the replacement of $\dvpK2$
by its Fourier component, the usage of \eq{lap} to replace the
Laplacian, similar to the first order case, and the application of
\eq{convolution} to terms quadratic in the first order field
fluctuation.

Rewriting the second part involving gradient terms, \eq{Fdvk1}, is
more involved, requiring the application of \eq{convolution} and the
use of \eq{gradients} as well as \eqs{lap} and (\ref{inv_lap}). We
leave the ``translation'' of individual terms in \eq{Fdvk1} to
Appendix \ref{app_fourier}, and give here immediately the result,
%
%
\bea
\label{Fdvk1_fourier}
&&
F\left(\dvpK1',\dvpK1\right)_\k
=\int d^3 \lambda 
\Bigg\{ 
\left(\frac{8\pi G}{\H}\right)^2\frac{p_kq^k}{q^2}
\delta\vp_{1I}'(\pvi)\sum_K\left(
X_K\dvpK1(\qvi)+\vp_{0K}'\dvpK1'(\qvi)\right)
\nonumber\\
&&
+p^2\frac{16\pi G}{\H}\dvpI1(\pvi)\sum_K\vp_{0K}'\dvpK1(\qvi)
+\left(\frac{4\pi G}{\H}\right)^2
\frac{\vp_{0I}'}{\H}\Bigg[
\left(p_lq^l-\frac{p^iq_jk^jk_i}{k^2}\right)\sum_K\vp_{0K}'\delta\vp_{1K}(\pvi)
\sum_K\vp_{0K}'\delta\vp_{1K}(\qvi)
\Bigg]\nonumber\\
&&
+2\frac{X_I}{\H}\left(\frac{4\pi G}{\H}\right)^2 
\frac{p_lq^lp_mq^m+p^2q^2}{k^2q^2}
\Bigg[
\sum_K\vp_{0K}'\delta\vp_{1K}(\pvi)
\sum_K\left(X_K\dvpK1(\qvi)+\vp_{0K}'\dvpK1'(\qvi)\right)
\Bigg]
\nonumber\\
&&
+\frac{4\pi G}{\H}
\Bigg[
4X_I\frac{q^2+p_lq^l}{k^2}\sum_K\left(
\dvpK1'(\pvi)\dvpK1(\qvi)\right)
-\vp_{0I}'p_lq^l\sum_K \delta\vp_{1K}(\pvi)\delta\vp_{1K}(\qvi)
\Bigg]
\nonumber\\
&&
+\left(\frac{4\pi G}{\H}\right)^2
\frac{\vp_{0I}'}{\H}\Bigg[
\frac{p_lq^lp_mq^m}{p^2q^2}
\sum_K\left( X_K\dvpK1(\pvi)+\vp_{0K}'\dvpK1'(\pvi)\right)
\sum_K\left(X_K\dvpK1(\qvi)+\vp_{0K}'\dvpK1'(\qvi)\right)
\Bigg]\nonumber\\
&&
+\frac{\vp_{0I}'}{\H}
\Bigg[
8\pi G
\sum_K\left(\frac{p_lq^l+p^2}{k^2}q^2\dvpK1(\pvi)\dvpK1(\qvi)
-\frac{q^2+p_lq^l}{k^2}\dvpK1'(\pvi)\dvpK1'(\qvi)
\right)
\nonumber\\
&&\qquad\qquad\qquad
+\left(\frac{4\pi G}{\H}\right)^2
\frac{k^jk_i}{k^2}\Bigg(
2\frac{p^ip_j}{p^2}
\sum_K\left(X_K\dvpK1(\pvi)+\vp_{0K}'\dvpK1'(\pvi)\right)
\sum_K X_K\dvpK1(\qvi)
\Bigg)\Bigg]
\Bigg\}\,.
\eea
%

\section{Slow roll}
\label{sr_sec}

Substantial simplification can be achieved by employing the slow-roll
approximation. In particular it will allow us to rewrite the source
terms of the second order Klein-Gordon equation solely in terms of the
field fluctuations $\dvpI1$ themselves, whereas in \eq{flatKG2real}
these terms also includes time derivatives of the first order field
fluctuations.

\subsection{Background and first order}
\label{sr_sec1}

The slow roll approximation in the background is given by
%
%
\be
\label{SRdef}
\vp_{0I}''-\H\vp_{0I}'\simeq 0\,, 
\qquad \sum_K\frac{1}{2a^2}{\vp_{0K}'}^2 \ll U_0 \,,
\ee
leading to 
%
%
\be
\label{background_sr}
3\H\vp_{0I}'+a^2U_{,\vp_I}=0\,, \qquad \H^2=\frac{8 \pi G}{3}a^2 U_0\,.
\ee
Making use of \eqs{background_sr} we find that \eq{defXI} simplifies to
%
%
\be
X_I=0\,.
\ee
We introduce the slow-roll parameters \cite{Seery2005}
\be
\label{def_epsI}
\epsilon_I\equiv{\sqrt{4\pi G}}\left(\frac{\vp_{0I}'}{\H}\right)\,,
\ee
and
\be
\label{def_etaIJ}
\eta_{IJ}\equiv
\frac{a^2 U_{,\vp_I\vp_J}}{3\H^2}\,.
\ee

At first order we get the Klein-Gordon equation from
Eq.~(\ref{KG1_flat}), imposing slow-roll in the background,
\be
\label{KG1_sr}
\dvpI1''+2\H\dvpI1'-\nabla^2\dvpI1
+\sum_K \M_{IK} \dvpK1=0\,,
\ee
where we defined the mass matrix
\be
\label{def_MIK}
\M_{IK}
\equiv 
a^2 U_{,\vp_I\vp_K}
-{24 \pi G}\vp_{0I}'\vp_{0K}'\,.
\ee
%
%
%
Note that $\eta_{IJ}$ is only small in the adiabatic direction, and
hence we keep terms including $\eta_{IJ}$ below.


\subsection{Second order}
\label{sr_sec2}

In the following we shall work to order $\epsilon_I^2$, where the slow
parameter $\epsilon_I$ is defined in \eq{def_epsI}.
We then get the Klein-Gordon equation at second order using slow roll
and working in Fourier space from \eqs{flatKG2real} and
(\ref{Fdvk1_fourier}),
%
%
\bea
\label{KG2_fourier_sr}
&&\dvpI2''(\kvi)+2\H\dvpI2'(\kvi)+k^2\dvpI2(\kvi)
+\sum_K\left(a^2
U_{,\vp_K\vp_I}-{24 \pi G}\vp_{0I}'\vp_{0K}'\right)
\dvpK2(\kvi) \\
&&+\int d^3 \lambda\ \Bigg\{
a^2\sum_{K,L} 
\left(
U_{,\vp_I\vp_K\vp_L} 
+ \frac{8\pi G}{\H}\vp_{0I}' U_{,\vp_K\vp_L}\right)
 \dvpK1(\pvi)\dvpL1(\qvi)
+\frac{16\pi G}{\H}a^2
\sum_K\vp_{0K}'\dvpK1(\pvi)\sum_K U_{,\vp_K\vp_I}\dvpK1(\qvi)\Bigg\}
\nonumber \\
&&+ \frac{8\pi G}{\H}
\int d^3 \lambda  \Bigg\{
\frac{8\pi G}{\H}\frac{p_l q^l}{q^2}\dvpI1'(\pvi)
\sum_K \vp_{0K}'\dvpK1'(\qvi)
+2p^2\dvpI1(\pvi)\sum_K \vp_{0K}'\dvpK1(\qvi)\nonumber\\
&&\qquad\qquad\qquad\qquad
+\vp_{0I}'
\sum_K\Bigg(
\left(\frac{p_lq^l+p^2}{k^2}q^2-\frac{p_lq^l}{2}\right)
\dvpK1(\pvi)\dvpK1(\qvi)
+\left(\frac{1}{2}-\frac{q^2+p_lq^l}{k^2}\right)
\dvpK1'(\pvi)\dvpK1'(\qvi)\Bigg)
\Bigg\}=0 \,.\nonumber
\eea
As expected, the simplification of the second order Klein-Gordon
equation in the slow roll limit, \eq{KG2_fourier_sr} above, compared
to the non-approximate versions given in previous sections,
\eqs{flatKG2real} and (\ref{Fdvk1_fourier}), is considerable.
%

\section{Discussion and conclusion}

We have derived the Klein-Gordon equation at second order in the
multi-field case in closed form in terms of the field fluctuations in
the flat gauge. We have given the full equation in real and Fourier
space, and also have given a simplified version where we used slow-roll.
Having calculated the Klein-Gordon equation at second order in closed
form for multiple fields in Section \ref{kg_sec_real} allowed us to
show that the Klein-Gordon equation also at this order does not
contain terms that diverge in the large scale limit, irrespective of
imposing the slow-roll approximation.
Preliminary calculations indicate, that this is the case as well if we
include vector perturbations.

If we want to solve the second order Klein-Gordon equation, we have to
proceed iteratively: first solving the background equations for the
fields $\vp_{0I}$ and the Hubble parameter $\H$, \eqs{KGback} and
(\eq{Friedmann}). Using these solutions we can then solve the first
order Klein-Gordon equation (\ref{KG1_flat}) and substitute the
results finally into the second order equation, (\ref{flatKG2real}).
This procedure is already implemented implicitly in the slow-roll case
discussed in Section \ref{sr_sec}, using the suitable slow-roll
expressions. However, if we are interested in numerical solutions,
this iterative scheme has to be explicitly taken into account. The
numerical implementation is however straight forward, and despite its
complexity, second order perturbation theory was shown to be faster
numerically than the $\Delta N$-formalism \cite{ML2006}.

However interesting or even exciting we may find the second order
Klein-Gordon equation to be, our aim is more likely to calculate what
observable consequences the evolution of the fields will have and what
it can tell us about our early universe model, and not the scalar
fields themselves. However, once the field dynamics is known, it is
easy use expressions relating the field fluctuations to observable
quantities. A particularly popular such quantity is the curvature
perturbation on uniform density hypersurfaces, $\zeta$, at first and
second order.
How in particular $\zeta_2$ is related to the field fluctuations on
flat slices, $\dvpI1$ and $\dvpI2$, used in this article, is given in
Ref.~\cite{M2005}, and its relation to the non-linearity parameter
$f_{\rm{nl}}$ is detailed in Ref.~\cite{ML2006}.

\acknowledgments

The author is grateful to David Burton, David Lyth, David Matravers,
David Seery, and David Wands for useful discussions and comments.  KAM
was supported by PPARC grant PPA/G/S/2002/00098.
Algebraic computations of tensor components were performed using
the \textsc{GRTensorII} package for Maple.

\appendix

\section{Useful background relations}
\label{app_back}

The Friedmann and the $i-j$ equations in the background,
\eqs{Friedmann} and (\ref{ij_back}), can be rewritten to give
\be
2\left(\frac{a'}{a}\right)^2-\frac{a''}{a}
=4\pi G \sum_K{\vp'_{0K}}^2\,.
\ee
Another useful combination is
\be
\H'+2\H^2
=8\pi G a^2 U_0\,.
\ee
%

\section{Klein-Gordon equation at second order}
\label{KG2_intermediate}

It might prove useful to have an intermediate version of the
Klein-Gordon equation at second order, that still has the metric
potentials at first order, whereas the second order metric potentials
and the time derivatives of the first order variables have been
replaced using the field equations given in Section \ref{field_sect}.
%
%
%
%
\bea
\label{flatKG2_intermediate}
\dvpI2''&+&2\H\dvpI2'-\nabla^2\dvpI2+a^2\sum_K U_{,\vp_I\vp_K}\dvpK2
+a^2\sum_{K,L} U_{,\vp_I\vp_K\vp_L}\dvpK1\dvpL1
+4\phi_1\left(\sum_K a^2 U_{,\vp_I\vp_K}\dvpK1-\nabla^2\dvpI1\right)
\nonumber\\
&+&\frac{8\pi G}{\H}\left[\vp_{0I}' a^2\delta U_2
+X_I \sum_K \vp_{0K}'\dvpK2
\right]
-4\delta\vp_{1I,k}'B_{1,}^{~~k}+4a^2 U_{,\vp_I}\phi_1^2
+\frac{16\pi G}{\H}\dvpI1'\Xkdvk
\nonumber\\
&+&\frac{8\pi G}{\H}\vp_{0I}'\left\{
2\phi_1\left(a^2\delta U_1+\Xkdvk\right)
+\frac{1}{2}\sum_K\left({\dvpK1'}^2
+\delta\vp_{1K,l}\delta\vp_{1K,}^{~~~~l}\right)
\right\}\nonumber\\
&+&2\frac{X_I}{\H}\,\nabla^{-2}
\left\{
\nabla^2 B_1\nabla^2\phi_1
-\phi_{1,kl}B_{1,}^{~~kl}
+8\pi G\sum_K\Big(\dvpK1'\nabla^2\dvpK1
+\delta\vp_{1K,l}'\delta\vp_{1K,}^{~~~~l}\Big)
\right\}\\
&+&\frac{\vp_{0I}'}{\H}\left\{
B_{1,kl}B_{1,}^{~~kl}-\phi_{1,k}\phi_{1,}^{~~k}
-\left(\nabla^2B_1\right)^2
-\left(\frac{4\pi G}{\H}\right)^2\left[
\left(\Xkdvk\right)^2-\left(\sum_K\vp_{0K}'\dvpK1'\right)^2\right]\right\}
\nonumber\\
&-&\frac{\vp_{0I}'}{\H}
\; \nabla^{-2}\Bigg\{
8\pi G\sum_K\Big(
\delta\vp_{1K,l}\nabla^{2}\delta\vp_{1K,}^{~~~~l}
+\nabla^{2}\dvpK1\nabla^{2}\dvpK1
+\dvpK1'\nabla^{2}\dvpK1'
+\delta\vp_{1K,l}'\delta\vp_{1K,}^{\prime~~~l}\Big)
\nonumber\\
&&\qquad\qquad\qquad
+\Big(
\frac{8\pi G}{\H}B_{1,~j}^{~~i}\, \Xkdvk-\phi_{1,}^{~~i}\phi_{1,j}
\Big)_{,i}^{~j}
\Bigg\}
=0\,.\nonumber
\eea
%

\section{Tensor components}
\label{app_tensor}

\subsection{The metric tensor}
\label{app_met}

The metric tensor up to second order for scalar perturbations is
\bea
\label{metric1app}
g_{00}&=&-a^2\left(1+2\phi_1+\phi_2\right) \,, \\
g_{0i}&=&a^2\left(B_1+\frac{1}{2}B_2\right)_{,i}\,, \\
g_{ij}&=&a^2\left[\left(1-2\psi_1-\psi_2\right)\delta_{ij}
+2E_{1,ij}+E_{2,ij}\right]\,.
\eea
and its contravariant form is
\bea
\label{metric2app}
g^{00}&=&-a^{-2}\left[1-2\phi_1-\phi_2+4\phi_1^2-
B_{1,k}B_{1,}^{~k}\right] \,, \\
g^{0i}&=&a^{-2}\left[B_{1,}^{~i}+\frac{1}{2}B_{2,}^{~i}
-2B_{1,k}E_{1,}^{~ki}+2\left(\psi_1-\phi_1\right)B_{1,}^{~i}
\right]\,, \\
g^{ij}&=&a^{-2}\left[\left(1+2\psi_1+\psi_2+4\psi_1^2\right)\delta^{ij}
-\left(
2E_{1,}^{~ij}+E_{2,}^{~ij}-4E_{1,}^{~ik}E_{1,k}^{~~j}
+8\psi_1E_{1,}^{~ij}+B_{1,}^{~i}B_{1,}^{~j}\right)
\right]\,.
\eea
Note, \eqs{metric1app} and (\ref{metric2app}) are without gauge
restrictions, i.e.~no gauge has been specified.

\subsection{Energy-momentum tensor in the flat gauge}
\label{Tmunu_sect}


The energy-momentum tensor for $N$ scalar fields with potential
$U(\vp_I)$ is then split into background, first, and second order
perturbations, using \eq{tensor_split}, as
\be
T^\mu_{~\nu}\equiv T^\mu_{(0)\nu} +\delta T^\mu_{(1)\nu}
+\frac{1}{2} \delta T^\mu_{(2)\nu}\,,
\ee
and we get for the components, from \eq{multiTmunu}, at zeroth order
\bea
T^{0}_{(0)0} &=& -\left(\sum_K\frac{1}{2a^2}{\vp'_{0K}}^2+U_0\right)\,, 
\qquad
T^{i}_{(0)j} =
\left(\frac{1}{2a^2}{\sum_K\vp'_{0K}}^2-U_0\right)\delta^i_{~j}\,,
\eea
at first order
\bea
\label{deltaT100}
\delta T^{0}_{(1)0} &=& -\frac{1}{a^2}\sum_K\left(
\vp_{0K}'\dvpK1'-{\vp'_{0K}}^2\phi_1
\right)-\dU1   \,, \\
\label{deltaT10i}
\delta T^{0}_{(1)i} 
&=& -\frac{1}{a^2}\sum_K\left(\vp_{0K}'\dvpK1_{,i}\right)\,, \\
\delta T^{i}_{(1)j} &=& \frac{1}{a^2}\left[
\sum_K\left(\vp_{0K}'\dvpK1'-{\vp'_{0K}}^2\phi_1\right)-a^2\dU1
\right]\delta^i_{~j}\,,
\eea
and at second order in the perturbations 
\bea
\delta T^{0}_{(2)0} &=& -\frac{1}{a^2}\sum_K\Big[
\vp_{0K}'\dvpK2'-4\vp_{0K}'\phi_1\dvpK1'-{\vp'_{0K}}^2\phi_2
+4{\vp'_{0K}}^2\phi_1^2+\dvpK1'^2+a^2\dU2 \nonumber\\
&&\qquad+\dvpdvpKll-{\vp_{0K}'}^2\BkBk \Big]
\,, \\
\label{deltaT20i}
\delta T^{0}_{(2)i} 
&=& -\frac{1}{a^2}\sum_K\left(\vp_{0K}'\dvpK2_{,i}
-4 \phi_1\vp_{0K}'\dvpK1_{,i}+2\dvpK1'\dvpK1_{,i}
\right)\,, \\
\delta T^{i}_{(2)j} 
&=& \frac{1}{a^2}\sum_K\Big[
\vp_{0K}'\dvpK2'-4\vp_{0K}'\phi_1\dvpK1'-{\vp'_{0K}}^2\phi_2
+4{\vp'_{0K}}^2\phi_1^2+\dvpK1'^2-\dvpdvpKll
-{\vp_{0K}'}^2\BkBk\nonumber\\
&&\qquad-2\vp_{0K}'\delta\vp_{1K,l}B_{1,}^{~~l}
-a^2\dU2\Big]\delta^i_{~j}
+ \frac{2}{a^2}\left(\vp_{0K}'B_1+\dvpK1
\right)_{,}^{~i}\delta\vp_{1K,j}\,.
\eea
%

%
%

\section{Relations in Fourier space}
\label{app_fourier}

%

Here we give the Fourier transform of some of the terms in \eq{Fdvk1}
(integrating out the $p^i$ dependence and omitting the $(2\pi)^{-3}$
factor:
\bea
\label{fourier_terms}
\left[{\dvp1}^2 \right]_k &=& \int d^3q\ \dvp1(q^i)\dvp1(k^i-q^i)\,, 
\nonumber \\
\left[\dvp1\nabla^2\dvp1 \right]_k 
&=& -\int d^3q\ (k-q)^2 \dvp1(q^i)\dvp1(k^i-q^i)\,,   \nonumber \\
\left[\delta\vp_{1,l}{\delta\vp_{1,}^{~~l}}\right]_k 
&=& - \int d^3q\   q_i\left(k^i-q^i\right) \dvp1(q^i)\dvp1(k^i-q^i)\,, 
\nonumber \\
\left[\delta\vp_{1,l}\nabla^{-2}{\delta\vp_{1,}^{~~l}}\right]_k 
&=&  \int d^3q\  q^{-2}{q_i\left(k^i-q^i\right)}
\dvp1(q^i)\dvp1(k^i-q^i) \,,
\nonumber \\
\left[\nabla^2\left(\delta\vp_{1,ij}\ppij\dvp1\right) \right]_k 
&=&  -k^2 \int d^3q\ (k-q)^{-4} q^i q^j  \left(k_i-q_i\right)
\left(k_j-q_j\right) \dvp1(q^i)\dvp1(k^i-q^i)   \,.
%
%
\eea
The terms of \eq{Fdvk1} not listed above are similar to the ones given
in \eq{fourier_terms} and can be readily derived using the expressions
above.

\section{Changes compared to earlier versions of the paper}
\label{changes_sec}

Here we give a list of changes in the second order governing equations
compared with the version of the paper originally published on the
archive, {\tt astro-ph/0610864v1}. This should make it easier to
verify that one is using the (most) correct version of the equations
\ldots However, we do not list mere ``cosmetic'' changes or
rearrangements in the equations.\\

\noindent
Changes in \emph{version 2}:
\begin{enumerate}
\item
\eq{Ein00_2}: wrongly set brackets corrected
\item
\eq{spatialtrace2}: missing $a^2$ added
\item
\eq{flatKG2real}: $\vp_{0I}'\frac{1}{3}\sum_K{\dvpK1'}^2$ in 2nd line 
removed, and  rewritten 
\item 
Imposing slow roll on first order Klein-Gordon equation removed
in Section \ref{sr_sec1} (only valid on large scales)
\item
\eq{KG2_fourier_sr}: second order Klein-Gordon equation equation in
slow roll limit completely rewritten in light of above changes
\item
\eq{flatKG2_intermediate}: $\frac{1}{3}\sum_K{\dvpK1'}^2$ in square bracket
in 2nd line removed
\end{enumerate}

\noindent
Changes in \emph{version 3} (corresponding to the published version):

Relaxing the requirement that the slow-roll parameter $\eta_{IJ}$ is
small (only needs to be small in the adiabatic direction) leads to
additional terms (of the form $\vp_0 U_{\vp\vp}\dvp1\dvp1$) in the slow
roll Klein Gordon equation \eq{KG2_fourier_sr}.\\

\noindent
Changes in \emph{version 4}:

Second order field equations corrected (\eqs{0i_2version2}, 
leading to changes in the second order
Klein-Gordon equation, \eqs{flatKG2_intermediate},
(\ref{flatKG2real}), (\ref{Fdvk1_fourier}), and
(\ref{KG2_fourier_sr}).\\

\noindent
Changes in \emph{version 5}:

Second order field equations, \eqs{ij2}, and (\ref{divdivij2})
corrected (the concept of an ``off-trace'' equation proved to be not
useful at second order), leading to changes in the second order
Klein-Gordon equation, \eqs{flatKG2_intermediate},
(\ref{flatKG2real}), (\ref{Fdvk1_fourier}), and
(\ref{KG2_fourier_sr}).
Background slow roll equations in conformal time, \eq{SRdef}, corrected
leading to changes throughout Section \ref{sr_sec}.
Also changes in definition of second order flat field perturbation, though
leading to no changes in this paper.

{}


\begin{thebibliography}{}



\bibitem{LLBook}
A.~R.~Liddle and D.~H.~Lyth,
\emph{Cosmological inflation and large-scale structure}, CUP,
Cambridge, UK (2000).



\bibitem{SB} D.~S.~Salopek and J.~R.~Bond,
Phys.\ Rev.\ D {\bf 42} (1990) 3936.


\bibitem{RS}
G.~I.~Rigopoulos and E.~P.~S.~Shellard,
[arXiv:astro-ph/0405185].
G.~I.~Rigopoulos, E.~P.~S.~Shellard and B.~W.~van Tent,
arXiv:astro-ph/0410486.
G.~I.~Rigopoulos, E.~P.~S.~Shellard and B.~J.~W.~van Tent,
arXiv:astro-ph/0504508.


\bibitem{SaTa}
M.~Sasaki and T.~Tanaka,
Prog.\ Theor.\ Phys.\  {\bf 99}, 763 (1998)
[arXiv:gr-qc/9801017].


\bibitem{LMS}
D.~H.~Lyth, K.~A.~Malik and M.~Sasaki,
JCAP {\bf 0505}, 004 (2005)
[arXiv:astro-ph/0411220].


\bibitem{Lyth:2005fi}
D.~H.~Lyth and Y.~Rodriguez,
arXiv:astro-ph/0504045.  


\bibitem{LV}
D.~Langlois and F.~Vernizzi,
Phys.\ Rev.\ Lett.\  {\bf 95}, 091303 (2005)
[arXiv:astro-ph/0503416];
%
D.~Langlois and F.~Vernizzi,
arXiv:astro-ph/0610064.







\bibitem{SaSt95}
M.~Sasaki and E.~D.~Stewart,
``A General analytic formula for the spectral index of the density
perturbations produced during inflation,''
Prog.\ Theor.\ Phys.\  {\bf 95}, 71 (1996)
[arXiv:astro-ph/9507001].


\bibitem{Starobinsky}
A.~A.~Starobinsky,
Phys.\ Lett.\ B {\bf 117}, 175 (1982);
A.~A.~Starobinsky,
JETP Lett.\  {\bf 42}, 152 (1985)
[Pisma Zh.\ Eksp.\ Teor.\ Fiz.\  {\bf 42}, 124 (1985)].




\bibitem{Bardeen80} 
J.~M.~Bardeen,
Phys.\ Rev.\ D {\bf 22}, 1882 (1980).


\bibitem{Mukhanov}
V.~F.~Mukhanov, L.~R.~W.~Abramo and R.~H.~Brandenberger,
Phys.\ Rev.\ Lett.\  {\bf 78}, 1624 (1997)
[arXiv:gr-qc/9609026].

\bibitem{Bruni}
M.~Bruni, S.~Matarrese, S.~Mollerach and S.~Sonego,
Class.\ Quant.\ Grav.\  {\bf 14}, 2585 (1997)
[arXiv:gr-qc/9609040].

\bibitem{Acquaviva}
V.~Acquaviva, N.~Bartolo, S.~Matarrese and A.~Riotto,
Nucl.\ Phys.\ B {\bf 667}, 119 (2003)
[arXiv:astro-ph/0209156].


\bibitem{Maldacena}
J.~Maldacena,
JHEP {\bf 0305}, 013 (2003)
[arXiv:astro-ph/0210603].


\bibitem{Nakamura}
K.~Nakamura,
Prog.\ Theor.\ Phys.\  {\bf 110}, 723 (2003)
[arXiv:gr-qc/0303090].


\bibitem{Noh}
H.~Noh and J.~c.~Hwang,
Phys.\ Rev.\ D {\bf 69}, 104011 (2004).


\bibitem{Bartolo:2001cw}
N.~Bartolo, S.~Matarrese and A.~Riotto,
Phys.\ Rev.\ D {\bf 65}, 103505 (2002)
[arXiv:hep-ph/0112261].


\bibitem{Rigopoulos:2002mc}
G.~Rigopoulos,
Class.\ Quant.\ Grav.\  {\bf 21}, 1737 (2004)
[arXiv:astro-ph/0212141].


\bibitem{Bernardeau}
F.~Bernardeau and J.~P.~Uzan,
Phys.\ Rev.\ D {\bf 67}, 121301 (2003)
[arXiv:astro-ph/0209330];
F.~Bernardeau and J.~P.~Uzan,
Phys.\ Rev.\ D {\bf 66}, 103506 (2002)
[arXiv:hep-ph/0207295].



\bibitem{MW2004}
K.~A.~Malik and D.~Wands,
Class.\ Quant.\ Grav.\  {\bf 21}, L65 (2004)
[arXiv:astro-ph/0307055].


\bibitem{BartoloJCAP}
N.~Bartolo, S.~Matarrese and A.~Riotto,
JCAP {\bf 0401}, 003 (2004)
[arXiv:astro-ph/0309692].


\bibitem{Finelli:2003bp}
F.~Finelli, G.~Marozzi, G.~P.~Vacca and G.~Venturi,
Phys.\ Rev.\ D {\bf 69}, 123508 (2004)
[arXiv:gr-qc/0310086].



\bibitem{Bartolo:2004if}
N.~Bartolo, E.~Komatsu, S.~Matarrese and A.~Riotto,
Phys.\ Rept.\  {\bf 402}, 103 (2004)
[arXiv:astro-ph/0406398].



\bibitem{Enqvist:2004bk}
K.~Enqvist and A.~Vaihkonen,
JCAP {\bf 0409}, 006 (2004)
[arXiv:hep-ph/0405103].


\bibitem{filippo}
F.~Vernizzi,
Phys.\ Rev.\ D {\bf 71}, 061301 (2005)
[arXiv:astro-ph/0411463].


\bibitem{Tomita:2005et}
K.~Tomita,
Phys.\ Rev.\ D {\bf 71}, 083504 (2005)
[arXiv:astro-ph/0501663].


\bibitem{Lyth:2005du}
D.~H.~Lyth and Y.~Rodriguez,
Phys.\ Rev.\ D {\bf 71}, 123508 (2005)
[arXiv:astro-ph/0502578].


\bibitem{Seery2005}
D.~Seery and J.~E.~Lidsey,
JCAP {\bf 0509}, 011 (2005)
[arXiv:astro-ph/0506056].


\bibitem{M2005}
K.~A.~Malik,
JCAP {\bf 0511}, 005 (2005)
[arXiv:astro-ph/0506532].


\bibitem{Seery:2006}
D.~Seery, J.~E.~Lidsey and M.~S.~Sloth,
arXiv:astro-ph/0610210.



\bibitem{Sasaki1986}
M.~Sasaki,
Prog.\ Theor.\ Phys.\  {\bf 76}, 1036 (1986).


\bibitem{Mukhanov88}
V.~F.~Mukhanov,
Sov.\ Phys.\ JETP {\bf 67}, 1297 (1988)
[Zh.\ Eksp.\ Teor.\ Fiz.\  {\bf 94N7}, 1 (1988)].



\bibitem{Hwang}
J.~c.~Hwang,
arXiv:gr-qc/9608018.


\bibitem{Taruya} 
A.~Taruya and Y.~Nambu,
Phys.\ Lett.\ B {\bf 428}, 37 (1998) 
[arXiv:gr-qc/9709035].



\bibitem{ML2006}
K.~A.~Malik and D.~H.~Lyth,
JCAP {\bf 0609}, 008 (2006)
[arXiv:astro-ph/0604387].






\end{thebibliography}
\end{document}